# Electron Mobility in Monoclinic β-Ga$_2$O$_3$ – Effect of Plasmon-phonon Coupling, Anisotropy, and Confinement

Krishnendu Ghosh[†] and Uttam Singisetti[§]

Electrical Engineering Department, University at Buffalo, Buffalo, NY 14260, USA

[†]kghosh3@buffalo.edu, [§]uttamsin@buffalo.edu

**Abstract**

This work reports an investigation of electron transport in monoclinic β-Ga$_2$O$_3$ based on a combination of density functional perturbation theory based lattice dynamical computations, coupling calculation of lattice modes with collective plasmon oscillations and Boltzmann theory based transport calculations. The strong entanglement of the plasmon with the different longitudinal optical (LO) modes make the role LO-plasmon coupling crucial for transport. The electron density dependence of the electron mobility in β-Ga$_2$O$_3$ is studied in bulk material form and also in the form of two-dimensional electron gas. Under high electron density a bulk mobility of 182 cm$^2$/ V.s is predicted while in 2DEG form the corresponding mobility is about 418 cm$^2$/V.s when remote impurities are present at the interface and improves further as the remote impurity center moves away from the interface. The trend of the electron mobility shows promise for realizing high electron mobility in dopant isolated electron channels. The experimentally observed small anisotropy in mobility is traced through a transient Monte Carlo simulation. It is found that the anisotropy of the IR active phonon modes is responsible for giving rise to the anisotropy in low-field electron mobility.

## I.  INTRODUCTION

β-Ga$_2$O$_3$ has recently emerged as a promising wide-bandgap material for future power electronic and optoelectronic applications. There have been several experimental demonstrations on high-power MOSFETS [1-3], Schottky diodes [4-6], and deep UV photodetectors [7]. Well-developed crystal growth technology and conventional processing techniques make it a further strong candidate beside its competitors SiC and GaN. Accuracy in n-type doping and difficulty in achieving a p-type doping make electrons the primary charge carrier of interest. So studying electron transport in this material is crucial to engineer the electronic device operation. Theoretical investigations on electronic structure (both ground states and excited states) in this material have been performed several times [8-11]. Lattice dynamical calculations predicted thermal conductivity and elastic properties [12-15]. Hall measurements [16] are also performed to study temperature dependent mobility as well as field-dependent mobility was investigated to study crystal-orientation dependent of mobility. Recently, we reported low-field [17] transport calculations in this material from first principles. It was found that the low-field mobility is limited by electron scattering due to polar optical phonon (POP) modes. This clearly reflects an intrinsic mobility limit of the material where the room temperature mobility hovers between 110 cm$^2$/Vs and 140 cm$^2$/Vs based on the direction of the applied field. On one hand the anisotropy of the mobility requires a firm understanding to help device design and on the other side improving the mobility is crucial for efficient power electronic operation which demands a high on-state current. A traditional way of improving impurity controlled mobility is to use hetero-junctions to spatially isolate the dopants from the electrons. The intrinsic mobility could be potentially improved by enhancing the free carrier screening of the POP interaction. Our previous work [17] did not consider any free carrier screening while estimating mobility. A recent work [18] did consider such screening under a static Lindhard dielectric function but it did not account for any dynamic frequency-dependent effects. However, considering the dynamic effects on the screening is important to probe the scattering rates accurately. It becomes further interesting for β-$Ga_2O_3$ due to the presence of many LO modes as, for a given electron density, the plasmon mode will couple with the different LO modes differently (some of them will be screened while the other will be anti-screened).

In this work, we study some important aspects of the electron mobility in β-$Ga_2O_3$ with significant details on the physics of plasmon-phonon coupling, dynamic screening and anisotropy. Initially lattice dynamical calculations are performed using density functional perturbation theory (DFPT) under local density approximation (LDA). Plasmon-phonon coupling is thoroughly investigated in the long-wavelength limit incorporating frequency dependent dynamic screening model under Lyddane-Sachs-teller (LST) theory and the plasmon-pole approximation. Scattering rates mediated by the coupled modes is computed using the Fermi-Golden rule and then the Boltzmann transport equation is solved iteratively to estimate the mobility at room temperature. The electron concentration dependence of the mobility is studied thereby. The anisotropy of the electron transport is studied through an interesting picture of anisotropic polar phonon emission based on Monte Carlo simulations. The mobility of two-dimensional electron gas (2DEG) formed at a heterojunction is studied including the coupled mode scattering and remote impurity scattering.

## II. ELECTRON-LO PHONON COUPLING IN β-$Ga_2O_3$

Low-field electron transport in β-$Ga_2O_3$ is controlled by the interaction between electrons and polar optical phonons as was revealed in our previous work and also in other recent studies [17, 19]. The room-temperature intrinsic electron mobility is limited by the POP even at a moderate doping of $1.0\times10^{17}$ /cm$^3$ which clearly reflects the strong polar coupling between electrons and phonons. In our previous work this coupling, $g^v_{POP}(\boldsymbol{q})$, for a given phonon mode $v$ and wave-vector $\boldsymbol{q}$ was computed using the Vogl model [20],

$$g^v_{POP}(\boldsymbol{q}) = \frac{e^2}{\Omega\varepsilon_0}\sum_\kappa \left(\frac{\hbar}{2M_\kappa\omega_{qv}}\right)^{\frac{1}{2}}\frac{\boldsymbol{q}.\boldsymbol{Z}_\kappa.\boldsymbol{u}_{\kappa v}(\boldsymbol{q})}{\boldsymbol{q}.\boldsymbol{\varepsilon}_\infty.\boldsymbol{q}} \qquad (1)$$

Here, $\boldsymbol{Z}_\kappa$, $\omega_{qv}$, and $\boldsymbol{u}_{\kappa v}(\boldsymbol{q})$ are the Born-effective charge tensor, phonon eigen energies, and cell-normalized phonon displacement patterns for an atom $\kappa$ for the mode $(\boldsymbol{q},v)$ respectively. $\boldsymbol{\varepsilon}_\infty$ represents the high-frequency dielectric tensor elements that contain screening contribution from the valence band electrons. $M_\kappa$ s denote the atomic masses while $\Omega$ is the unit cell volume. $e, \hbar$, and $\varepsilon_0$ are the unit electronic charge, reduced Planck's constant, and vaccum permittivity respectively. $\boldsymbol{Z}_\kappa$, $\boldsymbol{\varepsilon}_\infty$, $\omega_{qv}$, and $\boldsymbol{u}_{\kappa v}(\boldsymbol{q})$ are calculated under density-functional perturbation theory (DFPT) [21] using Quantum ESPRESSO [22]. To obtain a $\boldsymbol{q}$-space fine-resolution in $g^v_{POP}(\boldsymbol{q})$, $\omega_{qv}$, and $\boldsymbol{u}_{\kappa v}(\boldsymbol{q})$ are interpolated using a Wannier-Fourier interpolation scheme [23, 24]. Note that the overlap of the electronic wave-functions is taken to be unity due to the long-range nature ($\boldsymbol{q} \to \boldsymbol{0}$) of the coupling.

An important feature that accounts for the strength of the POP coupling is the splitting of the longitudinal optical (LO) and transverse optical (TO) modes near the zone-center. We compute this by adding the non-analytic macroscopic polarization contribution [25] to the DFPT produced force-constants. The non-analytical force-constant elements [21] have a form

$$C^{\alpha\beta}_{\kappa\kappa'}(\boldsymbol{q}) = \frac{4\pi e^2}{\Omega}\frac{(\boldsymbol{q}\boldsymbol{Z}_\kappa)_\alpha(\boldsymbol{q}\boldsymbol{Z}_{\kappa'})_\beta}{\boldsymbol{q}.\boldsymbol{\varepsilon}_\infty.\boldsymbol{q}} \qquad (2)$$

where $\alpha, \beta$ are the Cartesian directions. Diagonalizing the overall dynamical matrix (force-constants scaled by reduced atomic masses, $\sqrt{M_\kappa M_{\kappa'}}$) yields the LO eigen values. Fig. 1 shows the conventional unit cell (visualized by Vesta [26]) of β-Ga$_2$O$_3$ along with the Cartesian direction convention used in this work. There are 12 IR active phonons that could be categorized into two types – A$_u$ modes that are polarized along the *y* direction and the B$_u$ modes that are polarized on the *x-z* plane. The LO-TO splitting data obtained for all the IR active modes for three different Cartesian directions of the phonon wave-vector are shown in our previous work [17] that reveals the anisotropy of the LO-TO splitting. For example, the B$_u^1$ mode has a high splitting for ***q*** along *z* direction but very low splitting for the same along the *x* direction.

The important point to understand from this electron-POP coupling calculation is that the screening of the coupling elements contains only the contribution from the valence electrons which is clearly an underestimation of the overall screening that contains contribution from other phonons and plasmon. Especially in β-Ga$_2$O$_3$, with 12 IR active phonon modes, the phonon contribution is expected to take a major role in shaping the frequency dependence of the dielectric tensor elements. In our previous work, we made a very simple attempt to include this effect by using a conventional Lyddane-Sachs-Teller (LST) relation applicable to systems with orthogonal polarization vectors. This is an over-simplification of the screening problem given the non-orthogonal displacement patterns of the different phonon modes. Moreover, at high doping screening contribution from free carriers become important as well. The coupling between plasmon and phonon need to be addressed under such situation to accurately predict the role of free carrier-screening.

### III. ELECTRON – LOPC IN β-Ga$_2$O$_3$

Plasmons are longitudinal vibrational modes of collective electrons. The vibrational energy of such oscillations in a bulk semiconductor under long-wavelength limit is given by [27]

$$\omega_P^2 = \frac{\hbar^2 n_s e^2}{m^* \varepsilon_\infty} \tag{3}$$

Here, $n_s$ is the conduction electron density and $m^*$ is the electron effective mass. So for practical doping levels ranging from $10^{17}$ /cm³ to $10^{19}$ /cm³, the plasmon energy varies from about 10 meV to 100 meV in β-Ga$_2$O$_3$ considering an isotropic effective mass of 0.3 and an average isotropic high-frequency dielectric constant of 4.3 as calculated from our DFT and DFPT calculations respectively. This energy range is same as the range where all the LO phonon energies lie. Hence for transport calculations and subsequent device applications LO-plasmon coupling (LOPC) is expected to play a vital role.

### A. The LOPC modes at arbitrary wave-vectors

Although the pure plasmon energy is fairly isotropic, the LOPC modes would be highly anisotropic due to the anisotropy of the LO modes. Hence we compute the LOPC modes for each wave-vector separately. For a given wave-vector $\boldsymbol{q}$, the pure LO modes $\omega_{LO}^i(\boldsymbol{q})$ are computed by diagonalizing the DFPT computed dynamical matrix at the Γ point after adding the macroscopic polarization (Eq. 2). The effective dielectric constant along the direction of $\boldsymbol{q}$ is formulated as from LO mode polarization and plasmon-pole approximation -

$$\varepsilon_\omega(\boldsymbol{q}) = \varepsilon_\infty \prod_{i=1-12} \frac{\left(\omega_i^{LO}(\boldsymbol{q})\right)^2 - \omega^2}{\left(\omega_i^{TO}\right)^2 - \omega^2} - \frac{\varepsilon_\infty \omega_P^2}{\omega^2} \quad (4)$$

$\omega_i^{TO}$ are the corresponding TO mode energies. The valence electron contribution to the dielectric elements is $\varepsilon_\infty$. The plasmon contribution is taken to be present along $\boldsymbol{q}$ since plasma oscillation is a longitudinal oscillation. Note that Eq. 4 already accounts for the displacement patterns of the modes since the $\omega_i^{LO}(\boldsymbol{q})$ is calculated explicitly for each $\boldsymbol{q}$. Hence comparing with the work of Schubert et. al. [15], the effect of the cosine terms ( Eq. 21(a-d) of [15] ) arising on the net dipole oscillation strength of the different modes are included in Eq. 4 of this work. So essentially, Eq. 4 of this work is equivalent to Eq. 21(a-d) of [15]. However, as for transport properties we are only interested on the dielectric tensor along a given $q$, we consider the LO-TO splitting only along that direction since the latter accounts for the lattice polarization along that **q**. The zeros of $\varepsilon_\omega(q)$ yield the LOPC mode energies. However, if one is interested in obtaining the LOPC patterns (which are not required for our transport calculations), then the knowledge on the other two transverse components of the dielectric tensor is also needed. While the method of Schubert et. al. [15] is valid at $q = 0$, a method of obtaining the patterns at an arbitrary $q$ is described in the Supplementary Information.

Fig. 2 (a-c) show the LOPC mode energies for varying electron concentrations ($n_s$) at three different wave-vectors. Fig. 2(a) shows 9 LOPC modes that possess pure $B_u$ symmetry since the wave-vector lies on the *x-z* plane and hence there is no coupling to the $A_u$ modes. Similarly, Fig. 2(b) shows 5 LOPC modes that possess pure $A_u$ symmetry since the wave-vector lies along the *y* direction. On the other hand, in Fig. 2(c) we see 13 LOPC modes which have mixed symmetry. The black dashed lines in Fig. 2(a-c) show the pure plasmon energy. At low $n_s$, modes that have much higher energy than that of the plasmon (say, the blue line in Fig. 2(a) or the cyan line in Fig. 2(b)) possess the uncoupled LO mode energy. However, at higher $n_s$, modes that are below the plasmon mode (say, the green lines in both Fig. 2(a) and 2(b)) possess the energy of the TO mode. The high energy plasmon could efficiently screen the macroscopic polarization thereby mitigating the splitting. This has a significant effect in scattering rates that cannot be ignored. In the case of $B_u$ symmetry, Fig. 2(a) considers the situation when the $q$ vector is along *z* direction. But modes that are polarized more along *x* direction couples less with plasmon and hence their energy remains more or less flat (like the red line on Fig. 2(a)). This is of course a result of the low-symmetry of monoclinic crystals.

**B. Plasmon and phonon content of the LOPC modes**

In order to calculate the electron scattering rates mediated by the LOPC modes we need to separate out the plasmon and phonon contents for each mode. This is because scattering with pure plasma does not effectively provide any average momentum relaxation for the ensemble of electrons rather it just renders exchange of momentum among electrons. In the following we follow a method originally proposed by Fischetti et. al. [29] for studying coupling between interface phonons from the dielectric and plasmons in the semiconductor. We compute the plasmon content in an LOPC mode $\nu$ as

$$\Lambda_\nu^P(\boldsymbol{q}) = \frac{\prod_i\left(\left(\omega_\nu^{LOPC}(\boldsymbol{q})\right)^2 - \left(\omega_i^{LO}(\boldsymbol{q})\right)^2\right)}{\prod_{i\neq \nu}\left(\left(\omega_\nu^{LOPC}(\boldsymbol{q})\right)^2 - \left(\omega_i^{LOPC}(\boldsymbol{q})\right)^2\right)} \tag{5}$$

$\omega_i^{LO}(\boldsymbol{q})$ are the pure LO modes obtained in absence of any plasmon. The total phonon content of the mode would be $1 - \Lambda_\nu^P(\boldsymbol{q})$.

The relative contribution of the individual phonon modes in a given LOPC mode could also computed under a similar technique –

$$R_\nu^{LOj}(\boldsymbol{q}) = \frac{\prod_i\left(\left(\omega_\nu^{LOPC}(\boldsymbol{q})\right)^2 - \left(\omega_i^{LOPC,\,-LOj}(\boldsymbol{q})\right)^2\right)}{\prod_{i\neq \nu}\left(\left(\omega_\nu^{LOPC}(\boldsymbol{q})\right)^2 - \left(\omega_i^{LOPC}(\boldsymbol{q})\right)^2\right)} \tag{6}$$

Here the $\omega_i^{LOPC,\,-LOj}(\boldsymbol{q})$ are the modes obtained by forming Eq. 4 without the response of the $j_{th}$ LO mode and then setting its determinant to zero. The net contribution of the $j_{th}$ LO mode in the LOPC mode $\nu$ would be

$$\Lambda_\nu^{LOj}(\boldsymbol{q}) = \frac{R_\nu^{LOj}(\boldsymbol{q})}{\sum_k R_\nu^{LOk}(\boldsymbol{q})}\left(1 - \Lambda_\nu^P(\boldsymbol{q})\right) \tag{7}$$

Two important sum rules are to be verified here. First one of them is trivial, $\sum_j \Lambda_\nu^{LOj}(\boldsymbol{q}) + \Lambda_\nu^P(\boldsymbol{q}) = 1$, which says the total contribution of the plasmon mode and the LO modes to a particular LOPC mode must be 1. The second one is $\sum_v \Lambda_v^{LOj}(\boldsymbol{q}) = 1$ and $\sum_v \Lambda_v^{P}(\boldsymbol{q}) = 1$ which says the sum of the relative contributions of any LO mode (or the plasmon mode) to all the LOPC modes must be 1. The sum rules are verified for all the electron densities considered in this work.

Fig. 3 (a) and 3(b) show the plasmon contents in the different LOPC modes as we sweep up the electron concentration for modes with B$_u$ and A$_u$ symmetry respectively. The strong entanglement among the different branches is a signature of strong coupling. The sum over all modes for the plasmon content ($\sum_v \Lambda_v^{P}(\boldsymbol{q})$) is shown by the dashed curve on Fig. 3(a). The color scheme of the different branches in Fig. 3(a-b) is consistent with that used in Fig. 2(a-b) and the same scheme is maintained in subsequent figures. With increasing $n_s$ the higher LOPC modes get more of the plasmon flavor and the lower modes retain back their LO flavor. B$_u$ modes that are polarized more along $x$ direction have a low plasmon content since the $\boldsymbol{q}$ vector in Fig. 3(a) is along $z$ direction.

## C. The dressed interaction

The undressed Frölich vertex for electron-LO phonon interaction has the form [30]–

$$|g_v^{LO}(\boldsymbol{q})|^2 = \frac{e^2}{2\Omega\varepsilon_0}\left[\frac{\omega_v^{LO}}{q^2}\left\{\frac{1}{\varepsilon_\infty} - \frac{1}{\varepsilon_{DC}}\right\}\right] \tag{8}$$

As we can see from the term within the bracket, the squared interaction strength is proportional to the difference of the two inverse dielectric constants – $\varepsilon_{DC}$ includes a 'full-response' of the LO mode while $\varepsilon_\infty$ takes the LO mode to be frozen. Under the same spirit we calculate the interaction elements for each LOPC mode. For a given LOPC mode, a pair of dielectric constants are evaluated for each LO mode $(LOj)$ – $\boldsymbol{\varepsilon}_{\omega_v^{LOPC}}^{+LOj}(\boldsymbol{q})$ that including the 'full-response' of that LO mode and $\boldsymbol{\varepsilon}_{\omega_v^{LOPC}}^{-LOj}(\boldsymbol{q})$ that keeps the LO mode frozen while all others active. Here we are interested only on the element of the dielectric tensor that is along the $\boldsymbol{q}$ vector (see subsection. III.A for details).

Note that this type of formulation of the interaction strength includes all the screening (and also anti-screening) contribution in the long-wavelength. Hence the dressed interaction term could be cast as –

$$\left|g_{LOPC}^{v,\ LOj}(\boldsymbol{q})\right|^2 = \frac{e^2}{2\Omega\varepsilon_0}\left[\frac{\omega_v^{LOPC}(q)}{q^2}\left\{\frac{1}{\varepsilon_{\omega_v^{LOPC}}^{-LOj}(\boldsymbol{q})} - \frac{1}{\varepsilon_{\omega_v^{LOPC}}^{+LOj}(\boldsymbol{q})}\right\}\Lambda_v^{LOj}(\boldsymbol{q})\right] \quad (9)$$

In terms of the actual calculations performed, $\varepsilon_{\omega_v^{LOPC}}^{-LOj}(\boldsymbol{q})$ is estimated from Eq. 4 with the summation running over all $i$ for $i \neq j$ and $\varepsilon_{\omega_v^{LOPC}}^{+LOj}(\boldsymbol{q})$ is also estimated from Eq. 4 with the condition on the summation that $\omega = 0$ for $i = j$. The anisotropic dependence of the interaction is completely taken care of under this manifold because all the dielectric elements are formed taking into account the displacement patterns of the modes.

Fig. 4 (a-b) show the dressed oscillator strengths ($K_v^{LOPC}$, the term within the curly braces in Eq.9) for $A_u$ and $B_u$ symmetry modes respectively with the contribution from each LO flavor being summed up, $|K_v^{LOPC}(\boldsymbol{q})|^2 = \sum_j \left|K_{LOPC}^{v,\ LOj}(\boldsymbol{q})\right|^2$. Just to avoid confusion, the color scheme of the different branches in Fig. 4(a-b) is consistent with that used in Fig. 2(a-b) The important message to convey from here is the effect of dynamic screening on the interaction strengths. To illustrate, let us consider the black, yellow, and purple curves (here we refer to them as LOPC3, LOPC4 and LOPC5 respectively) in Fig. 4(a). Under low electron density, all the modes are LO flavored. With increasing density, first the LOPC3 gets the plasmon character and starts anti-screening LOPC4 and hence around a density of $2\times10^{18}$/cm³ LOPC4 interaction starts rising. Gradually the LO character shifts from LOPC3 to LOPC4. Then LOPC4 starts anti-screening LOPC5 and the interaction strength of the latter starts growing. Next the plasmon character shifts from LOPC4 to LOPC5. Since LOPC5 is higher than LOPC4, the former starts screening the latter and hence the interaction strength LOPC4 goes down. Finally as the plasmon character shifts further high in energy LOPC5 also gets well screened and its scattering strength goes down as well. So in a material like β-Ga$_2$O$_3$ with several LO modes, as the plasmon character propagates from one mode to the other, the screening and anti-screening of the modes shift accordingly. Hence for a given electron density on mode could be screened while the other could be anti-screened and hence a generic trend of the electron scattering rates and hence the mobility with an increasing electron density is not expected. In the next Section while discussing electron mobility this issue is further explored.

### D. Plasmon damping

Plasmons cease to behave as collective excitations in the electron-hole pair continuum (EHC). This is going to influence the LOPC modes and the dynamic screening depending upon the magnitude of the wave-vector, **q**. This damping becomes particularly important for degenerate doping. We treat this effect (approximately) by turning off the plasmon mode in the EHC. The upper boundary of EHC is given by [27], $\omega^+(q) = \frac{\hbar^2 k_F q}{m^*} + \frac{\hbar^2 q^2}{2m^*}$, where $k_F$ is the Fermi wave-vector calculated at zero temperature. Due to isotropic conduction band minima, the upper boundary of EHC is taken to be isotropic in β-Ga$_2$O$_3$ and hence the plasmon damping is dependent only on the magnitude of **q**. In terms of the actual calculation performed, the plasmon mode is taken off from the dispersion and dynamic screening calculation whenever, $\omega_P < \omega^+(q)$. The data shown in Fig. 2, Fig. 3, and Fig. 4 do not take into account the effect of plasmon damping in order clearly convey the LOPC picture in β-Ga$_2$O$_3$. However, the transport calculations are done including the effects of the damping. It is to be noted that this damping is different from the finite temperature plasmon damping arising from electron-phonon interactions that give rise to the imaginary part of the dielectric function. The latter type of damping is not considered in this work because that would involve self-consistently finding out the electron relaxation rates and the imaginary part of the dielectric constant which the authors believe would be extremely challenging especially for Ga$_2$O$_3$ with so many phonon modes.

**E. Electron scattering mediated by the LOPC modes**

Having found out the interaction elements we are ready to compute the electron scattering mediated by the LOPC modes for electron wave-vectors $\boldsymbol{k}$. The scattering rate is computed under Fermi-Golden rule as –

$$S_{LOPC}(\boldsymbol{k}) = \sum_{j,v,\boldsymbol{q}} w_{\boldsymbol{q}} \left| g_{LOPC}^{v,\ LOj}(\boldsymbol{q}) \right|^2 \frac{1}{(E_{\boldsymbol{k}+\boldsymbol{q}} - E_{\boldsymbol{k}} \pm \omega_v^{LOPC}(\boldsymbol{q}) - i\delta)} \tag{10}$$

Here $E_{\boldsymbol{k}}$ denotes the electron energy at wave-vector $\boldsymbol{k}$ and $w_{\boldsymbol{q}}$ is the weight of the $\boldsymbol{q}$ point from the Brillouin zone sampling. $\delta$ is a small energy smearing chosen to be 10 meV in this work. We used an isotropic parabolic electron band which is a very good assumption for β-Ga$_2$O$_3$ near the Γ-point. However, the scattering rates are computed at each $\boldsymbol{k}$ point separately to probe the anisotropy in the later transport calculation. Fig. 5 (a) shows the electron scattering rate for two levels of electron concentrations. The scattering rate at a lower electron concentration is higher at lower electron energies. This is because of the anti-screening of the low energy phonon modes. On the other hand, at relatively higher electron concentrations the lower energy phonon modes are screened but the higher energy phonons are anti-screened. This suppresses the scattering rate at lower electron energies but boosts it up at higher energies. This has a significant role in deciding the electron mobility.

**IV. ELECTRON MOBILITY**

The electron mobility is calculated by solving the Boltzmann transport equation (BTE) using our in-house codes based on Rode's iterative scheme [31]. Scattering mechanisms included are the LOPC scattering and the ionized impurity scattering. The ionized impurity scattering rate is calculated using a simple Brooks-Herring model with an isotropic parabolic electronic band of effective mass 0.3m$_0$. The trend of mobility is studied with respect to varying electron concentrations for bulk β-Ga$_2$O$_3$ and for two-dimensional electron gas formed at heterojunctions like Al$_x$Ga$_{2-x}$O$_3$/Ga$_2$O$_3$.

## A. Iterative BTE solver

The iterative BTE solver goes beyond the relaxation time approximation as it takes care of the inelasticity of the scattering which is particularly important at low electron energies. The electronic distribution function, $f(k)$, is split in two parts as $f(\mathbf{k}) = f^0(k) + f'(k)\cos\theta$. While the equilibrium part $f^0(k)$ is the Fermi-Dirac function, the non-equilibrium part $f'(k)$ is evaluated self-consistently using the scattering information as discussed below. The angle $\theta$ is between the applied electric field and the electron wave-vector **k**. $f'(k)$ under electric field $\mathbf{F}$ is calculated as

$$f_{i+1}'(k) = \frac{\left(S_i^{IN}(k) - \frac{e\mathbf{F}}{\hbar}\cdot\nabla_{\mathbf{k}} f^0(k)\right)}{\left(S^{OUT}(k) + \frac{1}{\tau_{el}}\right)} \quad (11)$$

Here the subscripts denote the iteration number in the self-consistent scheme. $S^{OUT}(k)$ and $S^{in}(k)$ are the net out-scattering and net in-scattering terms respectively from an electronic wave-vector $k$. Note that, although a scalar notation of the wave-vector is used for the arguments $S^{OUT}$ and $S^{in}$ in Eq. 11, they are calculated along each three directions separately and hence Eq. 11 is for any given direction at a time. The anisotropic effects that are supposed to arise from the anisotropic electron-LOPC is inherent in the out-scattering and in-scattering terms while evaluating them using Fermi-Golden rule. However, anisotropy from the band-structure (which is very minimal in β-Ga$_2$O$_3$) is not incorporated in this formulation. $S^{in}(k)$ is dependent upon $f'(k)$ by

$$S^{in}(k) = \int X f'(k')[P_{\mathbf{k}'\to\mathbf{k}}(1 - f^0(k)) + P_{\mathbf{k}\to\mathbf{k}'} f^0(k)]\, d\mathbf{k}' \quad (12)$$

Here, $P_{k'\to k}$ is the Fermi-Golden transition rate and $X$ is the cosine of the angle between $\boldsymbol{k'}$ and $\boldsymbol{k}$. Eq. 11 and Eq. 12 form a self-consistent pair which is solved iteratively starting from the initial condition given by RTA, $f'_0(k') = -\frac{e\boldsymbol{F}}{\hbar} \cdot \nabla_{\boldsymbol{k}} f^0(k)$.

Fig. 5(b) shows the antisymmetric part ( $f'(k)$ ) of the distribution function after the convergence in the iteration has been achieved. The small discontinuities in the distribution functions are results of the onset of emission of the dominant LOPC modes. Under a low electron concentration such onset occurs at a lower energy because the lower LOPC modes are anti-screened while with increasing electron concentration the onset point shifts to higher energies since the anti-screening behavior shifts to higher LOPC modes. On the other hand the symmetric part of the distribution does not contribute to net drift mobility. The electron mobility is calculated from the anti-symmetric part of the distribution function as $\mu_n = \frac{\sum_k v(k) f'(k)(\cos\theta)^2}{\sum_k f^0(k)}$ where $v(\boldsymbol{k})$ is the group velocity of the electrons at a wave-vector $\boldsymbol{k}$ and $v(\boldsymbol{k})\cos\theta$ represents the drift velocity along the electric-field.

## B. Bulk mobility and anisotropy

The room temperature bulk mobility is calculated under two conditions – with ionized impurity scattering and without that. The electron concentration is taken to be same as the dopant concentration for the former case. The calculated mobility is shown in Fig. 6. For the case without any impurity scattering the mobility initially shows a decline which is attributed to the anti-screening of low energy LOPC modes and hence a stronger scattering strength. At higher doping the mobility increases with increasing doping due to strong screening of the LOPC modes. It is to be noted here that the screening in [18] is essentially static which is good under high electron densities when the plasmon energy is higher than all the LO energies and hence the dielectric constant can be represented by static limits like Thomas-Fermi model. However, at moderate electron densities like $10^{18}$ /cm3, some of the LO modes are screened while others are anti-screened which would have consequences on the electron transport and mobility. Using a dynamic model is crucial to capture the interplay of screening and anti-screening. As seen from [18], the phonon limited electron mobility shows an increase at moderate electron densities due to free-carrier screening which is correct under the static limit. This work augments that fact by adding the contribution of anti-screening. In the case when ionized impurity scattering is present (circles in Fig. 6), the enhancement of mobility at higher doping is negligible since the scattering due to impurities compensates the reduction of LOPC mediated scattering under strong screening. The mobility, including ionized impurity scattering, at an electron concentration of $5\times10^{19}$/cm$^3$ is about 182 cm$^2$/V.s. However for devices where the electronic channel is separated from the dopants an intermediate mobility (between the circles and triangles) is to be expected due to the presence of remote ionized impurity scattering which is discussed next during 2DEG mobility analysis.

Next we turn to discuss the anisotropy of the computed mobility in the two different directions. As seen from Fig. 6 the mobility is higher along the *y* direction compared to that in the *z* direction. The anisotropy is about 20% at a moderate doping. Such anisotropy is experimentally observed [32], but its origin is not clearly understood since it is well known that the electronic bands in β-Ga$_2$O$_3$ near the Γ point is isotropic. We attribute this anisotropy of the long-range interaction between the electrons and the LO modes (even in absence of any plasmon). The dominating B$_u^1$ mode as found in our previous work has a high projection of polarization along the *z* axis. Now low energy electrons moving along the *z* direction will get scattered by phonons with wave-vector along the *z* direction. This idea is shown on Fig. 7(a). To better convey the idea behind the origin of the anisotropy we carried out a Monte Carlo simulation to probe the emission rates of the different phonons mediated by the long-range interactions with electrons. No plasmon is considered in the MC simulation for simplicity and that does not affect the fundamental concept behind this anisotropy. Fig. 7(b) shows the emission rates of the three modes $B_u^1$, $A_u^2$, and $B_u^6$ under an external electric field of $5\times10^6$ V/m applied along the *z* direction. The emission rate of the $B_u^1$ is higher because of the stronger interaction. The $B_u^6$ mode has a relatively higher energy and only a few electrons have enough energy under this electric-field to emit $B_u^6$ modes. Now as the electric field is enhanced, as shown in Fig. 7(c), the emission rate of the $B_u^6$ mode increases because electrons gain enough energy. However, although $A_u^2$ has a lower energy than $B_u^6$, its emission rate increases by a smaller amount than $B_u^6$. This is because the for an applied electric field along the *z* direction the momentum of the electrons are more incline along the *z* direction and hence they couple more with phonons with wave-vectors along *z*. This explains the experimentally observed [32] anisotropy in electron mobility. Hence this anisotropy completely follows from the anisotropy of the long-range electron-phonon interaction and on contrary to most conventional semiconductors, this anisotropy is not a result of any conduction band anisotropy rather it is a clear signature of the low-symmetry of the monoclinic β-$Ga_2O_3$ crystal that results to anisotropic LO-TO splitting. The anisotropy decreases with increasing impurity scattering (see Fig. 6) since the latter is isotropic.

## C. 2DEG mobility

Two-dimensional electron gas in β-$Ga_2O_3$ has been very recently demonstrated experimentally [33]. Theoretical mobility limits are not well known yet. Here we study the mobility of 2DEG formed in the inversion layer of a simple $Al_xGa_{2-x}O_3$/$Ga_2O_3$ (ALGO/GO) heterojunction with varying electron concentration. The typical structure is shown in the inset of Fig. 8 where the 2DEG is situated at a distance *d* from the dopants. The dopants are taken to be as a sheet charge density behaving like a δ Coulomb potential and it is assumed that the 2DEG density at the channel is same as the sheet charge density of the dopants (such assumption could be relaxed by using a Poisson solver and is not addressed here for the sake of clearly conveying the trend of the mobility with increasing electron concentration). The scattering rate from such remote impurity (RI) center is modelled using a statically screened Coulomb interaction [34] –

$$S_{RI} = \frac{N_I e^4 m^*}{4\pi\hbar^3(\varepsilon_s+\varepsilon_s')^2} \int_0^{2\pi} \frac{(1-\cos\varphi)}{(q^2+q_{TF}^2)} A^2(q) \, d\theta \tag{13}$$

Here $N_I$ is the surface charge density present at the interface, $\varepsilon_s$ and $\varepsilon_s'$ are the static dielectric constants of the electron channel (in GO) and the dopant location (ALGO). We considered $\varepsilon_s = \varepsilon_s'$, which is not a bad approximation for small aluminum content in ALGO. $q_{TF}$ is the two dimensional Thomas-Fermi (TF) screening wave-vector given by [34] $\frac{2}{a_B^*}$, where $a_B^*$ is the effective Born radius. Note that he 2D TF wave-vector is independent of the electron concentration. $A(q)$ is the overlap function between the confined out-of-plane envelope function of the electron gas and the exponentially decaying Coulomb potential envelope from the remote impurities. The envelope function for the 2DEG is taken as the usual Fang-Howard form [34] with an average inversion layer thickness of 5 nm. The $(1-\cos\varphi)$ term in the numerator of Eq. 13 accounts for the effective momentum relaxation with $\varphi$ being the angle between the in-plane electron wave-vector (**k**) and **q**. Due to the elastic nature of the impurity scattering, $q = 2k\sin\frac{\varphi}{2}$.

The LOPC scattering formulation in the case of 2DEG follows the same steps as that for the bulk case except that the plasmon energy is modified by the first order term, of the plasmon dispersion for a 2DEG, $\omega_P^2 = \frac{\hbar^2 n_s e^2 q}{m^* \varepsilon_\infty}$, as shown by Stern [35]. The 2DEG is taken to be in the 1st sub-band and no LOPC mode mediated inter-subband transition is considered due to high enough energy gap with the second subband. The confined direction is taken to be the y direction. This implies that only the $B_u$ character LOPC modes are able to cause the scattering under this circumstance since momentum conservation will not allow the LOPC modes with wave-vector along the *y* direction to cause intra-subband transition. The computed mobility along the *z* direction is shown on Fig. 8 for several cases. Like the bulk case, the mobility improves with increased electron concentration due to enhanced screening of the LOPC modes. The anti-screening behavior is not observed since that occurs at a lower electron concentration than what is shown in Fig. 8. The mobility at a 2DEG density of $5\times10^{12}$ /cm$^2$, when the RI center is at the interface, is around 418 cm$^2$/V.s which is more than 2X higher than the bulk case. As the RI center moves away from the interface the mobility improves on the higher $n_s$ side due to less scattering by the RI. Reduction of the phase space for final state after scattering and the remoteness of the impurities are responsible for the improvement in mobility. The error bars in Fig. 8 are showing a $\pm10\%$ offset that might arise from issues like dielectric mismatch at the interface, truncating the plasmon dispersion after first-order, and any numerical inaccuracies. In reality this $\delta$ doping is only a few nanometers far from the interface in order to maximize the electron concentration in the channel. So as seen from Fig. 8 it is expected that the mobility would be close to 1000 cm$^2$/Vs in the absence of any other scattering mechanisms that could potentially originate from surface-roughness, alloy disorder, or remote interface phonons.

## V. CONCLUSION

We have calculated the electron density dependence of the mobility of β-Ga$_2$O$_3$ in bulk and 2DEG form. The enhanced screening at higher electron densities provide promise for improved mobility which is important for device operation. The interplay of screening and anti-screening of the LOPC modes at intermediate electron densities gives rise to interesting trends in the electron mobility. The anisotropy of the electron mobility is explained by an anisotropic polar phonon emission picture produced by Monte Carlo simulations. The 2DEG mobility shows more than 2X improvement than bulk mobility. Further study on the 2DEG mobility is required by changing the separation of the dopants and the 2DEG. Also the confinement direction can be changed from *y* to *z* for studying any further improvement of mobility.

The authors acknowledge the support from the National Science Foundation (NSF) grant (ECCS 1607833). The authors also acknowledge the excellent high performance computing cluster provided by the Center for Computational Research (CCR) at the University at Buffalo.

**Figures**

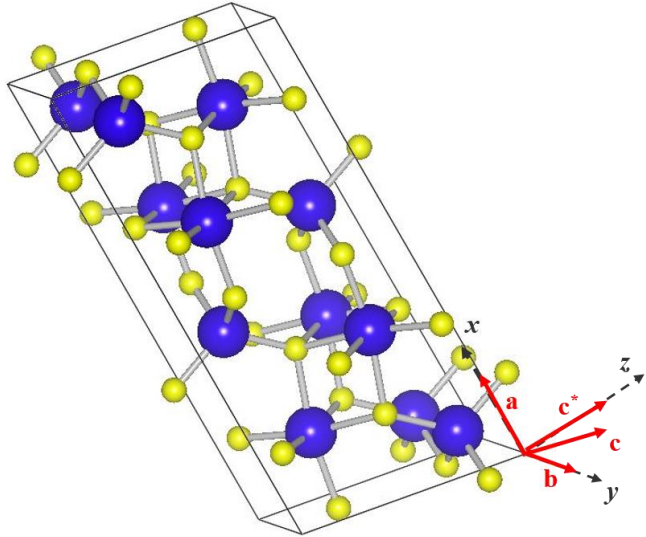

Fig 1: (Color online) The conventional unit cell of β-$Ga_2O_3$ visualized by Vesta [26]. Bigger atoms are Ga and smaller ones are O. The Cartesian direction convention is shown that is followed throughout this work.

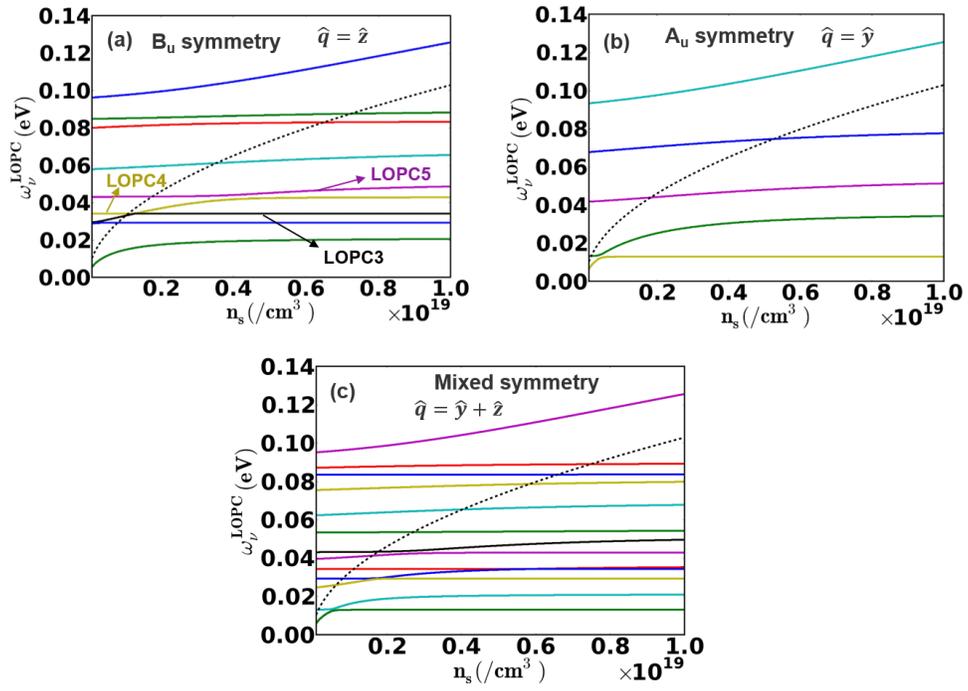

Fig. 2: (Color online) (a) The $B_u$ symmetry LOPC modes for the wave-vector along the $z$ direction. (b) The $A_u$ symmetry LOPC modes. (c) Mixed symmetry modes for the wave-vector along $\hat{y}+\hat{z}$. The pure plasmon modes are shown in dashed line for all the three cases.

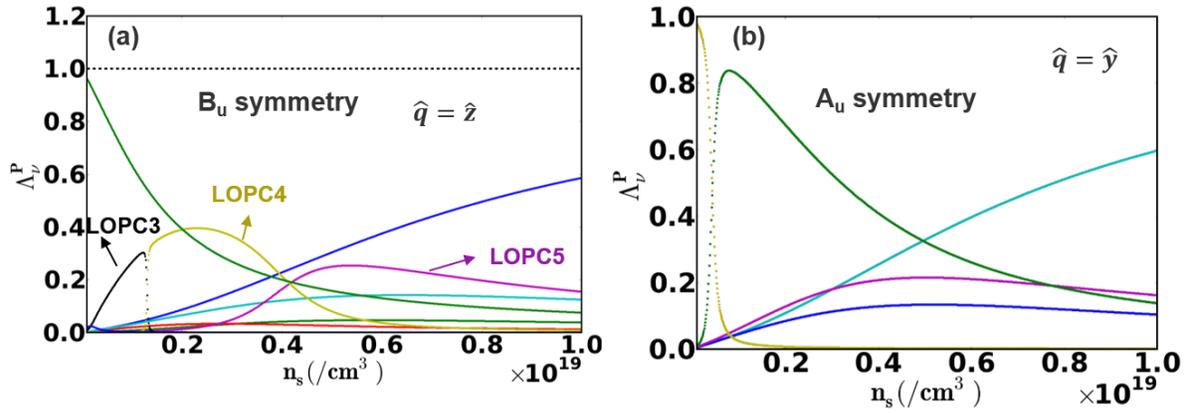

Fig. 3: (Color online) (a) The plasmon content of the $B_u$ symmetry LOPC modes, (b) similar plots for the $A_u$ modes. The strong entanglement reflects the possible influence of the plasmon in scattering strength. See text for details.

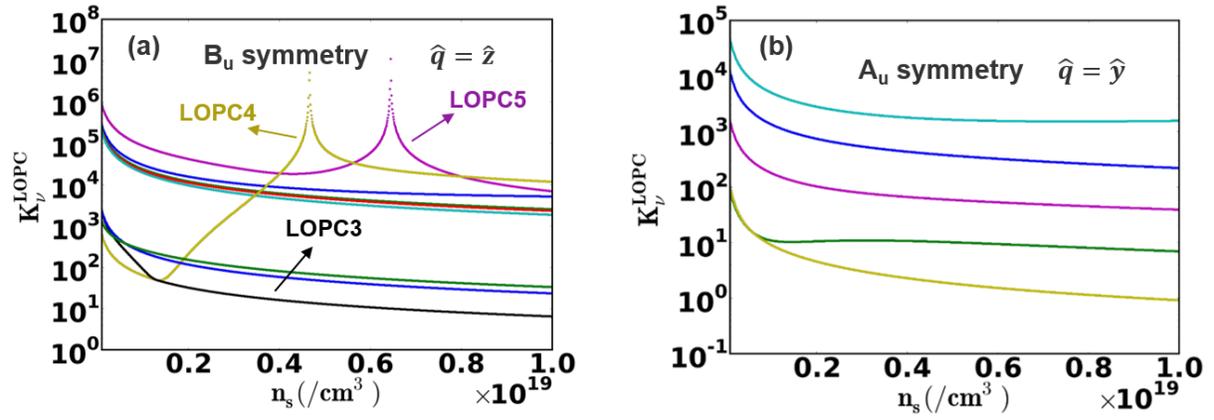

Fig. 4: (Color online) (a) The dynamically screened oscillation strength of the $B_u$ symmetry LOPC modes, (b) similar plots for the $A_u$ modes.

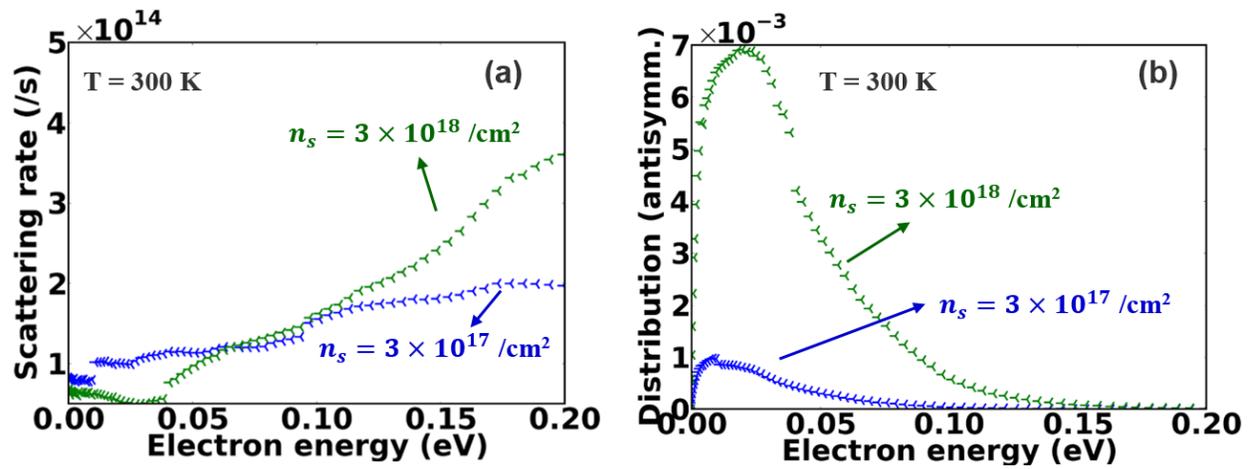

Fig. 5: (Color online) (a) The LOPC mediated electron scattering rates for two different levels of electron densities. (b) The electron distribution functions after the convergence of the iterative BTE scheme.

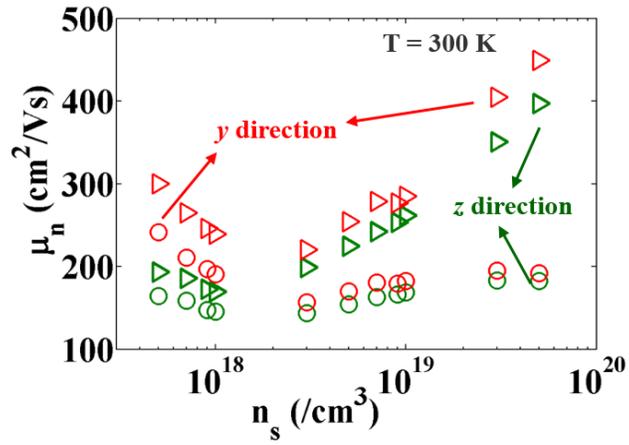

Fig. 6: (Color online) Electron density dependence of bulk mobility for two different Cartesian directions. The triangles show the mobility computed without any ionized impurity scattering while the circles show the mobility including ionized impurity scattering.

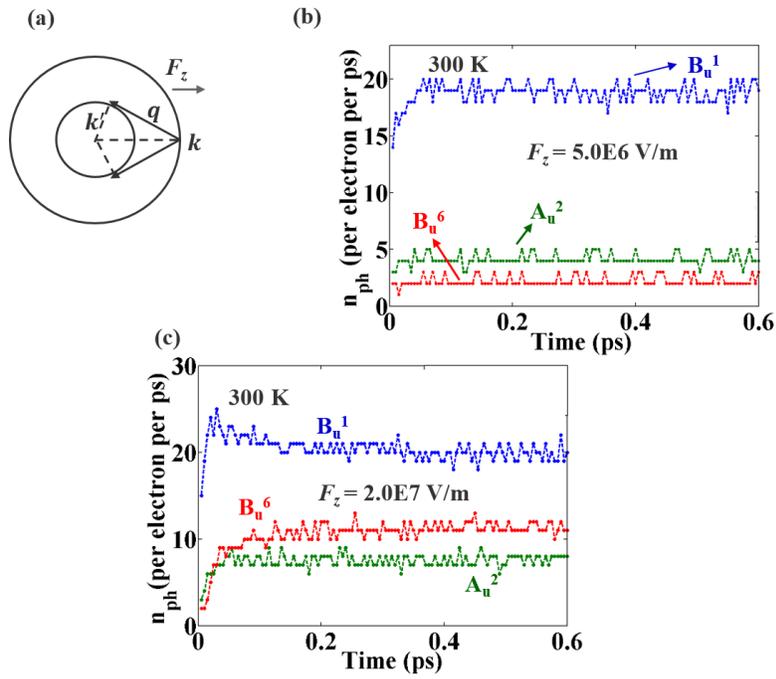

Fig. 7: (Color online) (a) The momentum conservation requires low energy electrons to be scattered by phonons whose wave-vectors are inclined towards (opposite) to the electron wave-vectors. (b) Emission rate of three IR active phonons mediated by long-range interaction with electrons under an external applied field of $5 \times 10^6$ V/m. (c) Same plots when the applied field is $2 \times 10^7$ V/m.

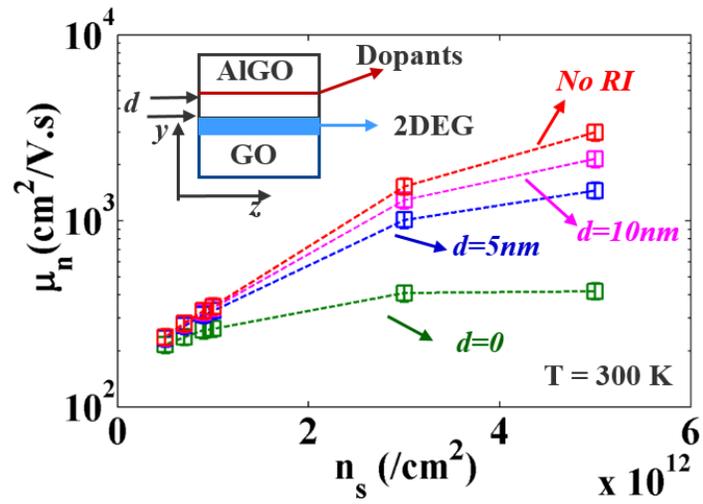

Fig. 8: (Color online) Electron density dependence of 2DEG mobility along $z$ direction. The mobility improves drastically as the RI center moves away from the interface. Error bars are included due to the approximations (see text for details). (Inset) shows the spatial location of the 2DEG and dopants.